\documentclass[doublecol]{epl2} 
\usepackage[utf8]{inputenc}
\usepackage{scalefnt}
\usepackage{url}

\title{Nonextensive entropic behavior observed in Quasar 3C 273}
\shorttitle{Nonextensive entropic behavior observed in Quasar 3C 273} 

\author{C. V. da Silva\inst{1} \and S. G. A. Barbosa\inst{1} \and F. V. Alencar Filho\inst{1} \and J. P. Bravo\inst{2} \and D. B. de Freitas\inst{1}}
\shortauthor{Barbosa et al.}

\institute{                    
  \inst{1} Departamento de F\'{\i}sica, Universidade Federal do Cear\'a, Caixa Postal 6030, Campus do Pici, 60455-900 Fortaleza, Cear\'a, Brazil\\
  \inst{2} Avignon Universite, Campus Jean-Henri Fabre Centre d'Enseignement et de Recherche en Informatique Agroparc BP 91228, 84911 Avignon Cedex 9, France
}
\pacs{nn.mm.xx}{First pacs description}
\pacs{nn.mm.xx}{Second pacs description}
\pacs{nn.mm.xx}{Third pacs description}

\abstract{
We investigate the flux intensities spanning from radio waves to X-rays across 39 light curves of Quasar 3C 273, utilizing publicly available data collected by the Integral Science Data Centre (ISDC) database. Our results suggest that Quasar 3C 273 exhibits nonextensive behavior. Furthermore, we calculate the $q$ entropic indices for these light curves using the $q$-Gaussian distribution with a predominant observation of cases where $q>1$. Based on this index, we estimate the non-extensive entropy ($S_{q}$) and explore its correlation with the energy (in eV). In this context, we identify two jump-like increases in entropy, particularly evident in the infrared (IR) and X-ray wavebands. The peak in the far-IR band, around 0.34 eV, results from synchrotron flares evolving from higher to lower energies and thermal radiation emitted by hot dust near the sublimation radius. However, the second entropic peak in the hard X-ray range lacks statistical robustness due to limited data or large measurement uncertainties.}

\begin{document}

\maketitle

\section{1) Introduction}
Active Galactic Nuclei (AGNs) are known for their extreme brightness and high variability across multiple wavelengths that extends from radio to $\gamma$-rays \cite{1979ApJ...232...34B,1995PASP..107..803U,galaxies11050096}. The Quasar 3C 273, a bright and nearby ($z$ = 0.158) radio-loud quasar, has been extensively studied across multiple frequencies for decades, making it one of the most well-documented AGNs \cite{1963Natur..197,1987A&A...176..197C,1990A&A...234...73C,2008MmSAI..79.1011C}. The emissions of AGNs, including 3C 273, are complex due to the presence of multiple emission components, such as synchrotron radiation and thermal emission. Several studies \cite{2008A&A...486..411S, 2006A&A...451L...1T,2000A&A...361..850T,1993MNRAS.262..249R,1988Natur.335..330C} point out that, in 3C 273, synchrotron flares from a relativistic jet dominate the radio-to-millimeter energy output and extend up to the IR and optical domains. Additionally, a UV source's thermal emission from dust grains is partly responsible for producing an IR continuum \cite{10.1088/0004-637x/753/1/33}.

According to Soldi \textit{et al.} \cite{2008A&A...486..411S}, the bright excess in the optical-UV band is interpreted as a signature of the accretion disc. However, Courvoisier \& Turler \cite{2005A&A...444..417C} suggest that clumping matter, rather than an accretion disc, may produce UV radiation.  The interaction of matter accreted in clumps produces optically thick shocks that emit UV light, while optically thin shocks closer to the black hole produce X-rays. Comptonization of hot plasma in an accretion disc can create X-ray emission comparable to that found in Seyfert galaxies \cite{2004Sci...306..998G}.  According to Page \textit{et al.}\cite{2004MNRAS.349...57P}, the soft excess may be caused by thermal Comptonization of cool-disc photons in a warm corona. On the other hand, Grandi \& Palumbo \cite{2004Sci...306..998G} and Kotaoka \textit{et al.} \cite{2002MNRAS.336..932K} suggest that the X-ray to $\gamma$-ray emission is thought to be caused by inverse Compton processes in a thermal plasma in the disc or corona, as well as a non-thermal plasma associated with the jet. 

The Integral Science Data Centre (hereafter ISDC), which hosts the light curves of 3C 273, is one of the most complete multiwavelength repositories available for an AGN. Along with the High-Energy Astrophysics Virtually ENlighted Sky (HEAVENS) web interface, the ISDC offers extensive observational data, including more than 40 years of light curves across various wavebands \cite{1999A&AS..134...89T,2008A&A...486..411S,2010int..workE.162W}. These datasets have enabled detailed studies of the variability of Quasar 3C 273, particularly at high energies, ranging from keV to GeV. Variability at these energies has been analyzed using instruments such as IBIS and SPI on board INTEGRAL, PCA on RXTE, and LAT on Fermi \cite{1996SPIE.2808...59J,2009ApJ...697.1071A}.

The variability of AGNs, including Quasar 3C 273, has been studied using various methods to quantify amplitude and periodicity, revealing complex emission behavior likely originating from different locations within the AGN \cite{2008A&A...486..411S}. Methods like the discrete autocorrelation function, the Lomb-Scargle periodogram, and the wavelet transform have been employed to study the quasi-periodic and periodic variability of AGN sources and their correlations with several statistical parameters, such as standard deviation of signal, thermodynamical properties and short- and long-term fluctuations \cite{1998ApJ...503..662H,2006A&A...456L...1K,2007A&A...469..899H,2006AdSpR..38.1405V,2016MNRAS.461.3145V,2015Natur.518...74G}. 

There is still a lack of studies investigating quasars as sources of energy in out-of-thermodynamical equilibrium. 
An important work in this research line is Rosa \textit{et al.} \cite{ROSA20136079}'s study, which emphasizes the role of non-extensive statistical mechanics in investigating the entropic behavior of various astrophysical sources from Novae to Pulsars. However, no studies have yet analyzed the entropic behavior of quasars across different wavelengths. To fill this gap, non-equilibrium statistical approaches, particularly non-extensive statistical mechanics, provide a valuable framework for understanding how self-gravitating systems manage energy retention and release mechanisms \cite{tsallis1988,Tsallis4}. 

Our paper analyzes the Quasar 3C 273 from radio waves to $\gamma$-ray as a non-extensive phenomenon. In addition, we study their degree of nonextensivity, analyzing the distributions of 39 light curves. Our main aim is to investigate the entropic behavior as a function of the energy of the source. The remainder of this paper is organized as follows: Section 2 describes the data and procedures used. In Sect. 3, we report the results and discussions, including a comparison with the literature. The final remarks are presented in the last section.

\section{Working sample}\label{data}
We selected all the 3C 273 public data from 1963 to 2005 in the ISDC database\footnote{\url{http://isdc.unige.ch/3c273/}} (see Table \ref{tabobs}), following the criteria outlined by Belete \textit{et al.} \cite{2018MNRAS.478.3976B}. The database is organized into 70 ASCII files, each containing a light curve for a specific wavelength. Each file contains ten columns, including the date in decimal year, frequency, wavelength, flux, flux error, and reference, with one row representing each observation. 

Although the ISDC database provides access to many multiwavelength light curves from radio to $\gamma$-ray ranges, we selected those with sufficient data over an extended period to study the long-term variability of this AGN \cite{2008A&A...486..411S,1999A&AS..134...89T}. Therefore, several wavelengths are selected for each spectral emission covering the following waveband: radio from 5 to 37 GHz, millimeter and submillimeter from 0.8 to 3.3 mm, infrared from 1.25 to 3.6 $\mu$m, optical from 3439 to 7000 \AA, ultraviolet from 1300 to 3000 \AA, and X-ray from 5 to 200 KeV. 

The selection criteria for data from the ISDC database are based on data quality and amount of data. To ensure data reliability and minimize noise in our analysis, we implemented a flagging system to categorize the quality of each data point. The Flag column can take integer values ranging from -3 to +1, with a default value of zero assigned to good data points, and negative flags indicating less reliable data. A flag of -1 denotes low-quality data, consistent with contemporaneous observations but with larger uncertainties that add unnecessary noise to the light curves. A flag of -2 represents ``uncertain'' data, often with large uncertainties or processed automatically without manual verification. A flag of -3 indicates "dubious" data points, which are clear outliers, incompatible with contemporaneous observations or the overall variability trend. On the other hand, High-quality IR and optical data accompanied by synchrotron flares lasting several days to weeks receive a flag of +1 \cite{10.1086/422499}. We manually flagged the data through a careful inspection of the light curves. In our analysis, we only include data with flags of 0 or +1 to construct our sample of light curves. Our sample is exposed in Table \ref{tabobs}. The date range (between the first and the last observation) and the number of data points correspond to reliable data (\texttt{Flag}$\ge$0).

\begin{figure}
	\begin{center}
		\includegraphics[width=0.38\textwidth,trim={2.5cm 1.5cm 2.5cm 1.5cm}]{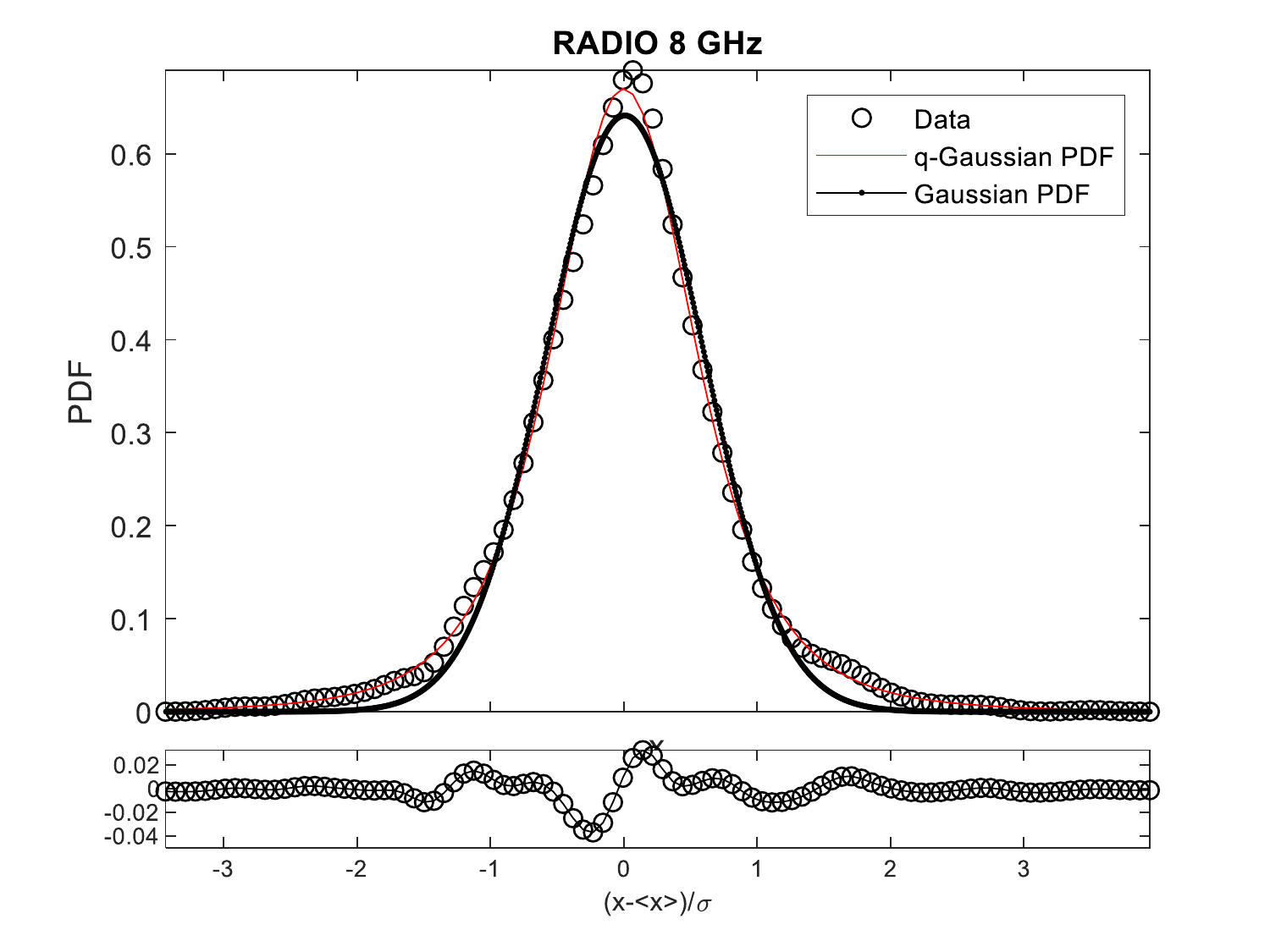}
	\end{center}
\caption{PDFs of flux for Radio 8 GHz light curve. The open circles represent data; the red and black curves fit with a $q$-Gaussian (Tsallis) distribution and standard Gaussian, respectively. The bottom panel denotes the residual between the empirical distribution and the $q$-Gaussian PDF.}
\label{FigqRadio8GHz}
\end{figure} 

\begin{figure}
	\begin{center}
		\includegraphics[width=0.49\textwidth,trim={2.5cm 1.5cm 2.5cm 1.5cm}]{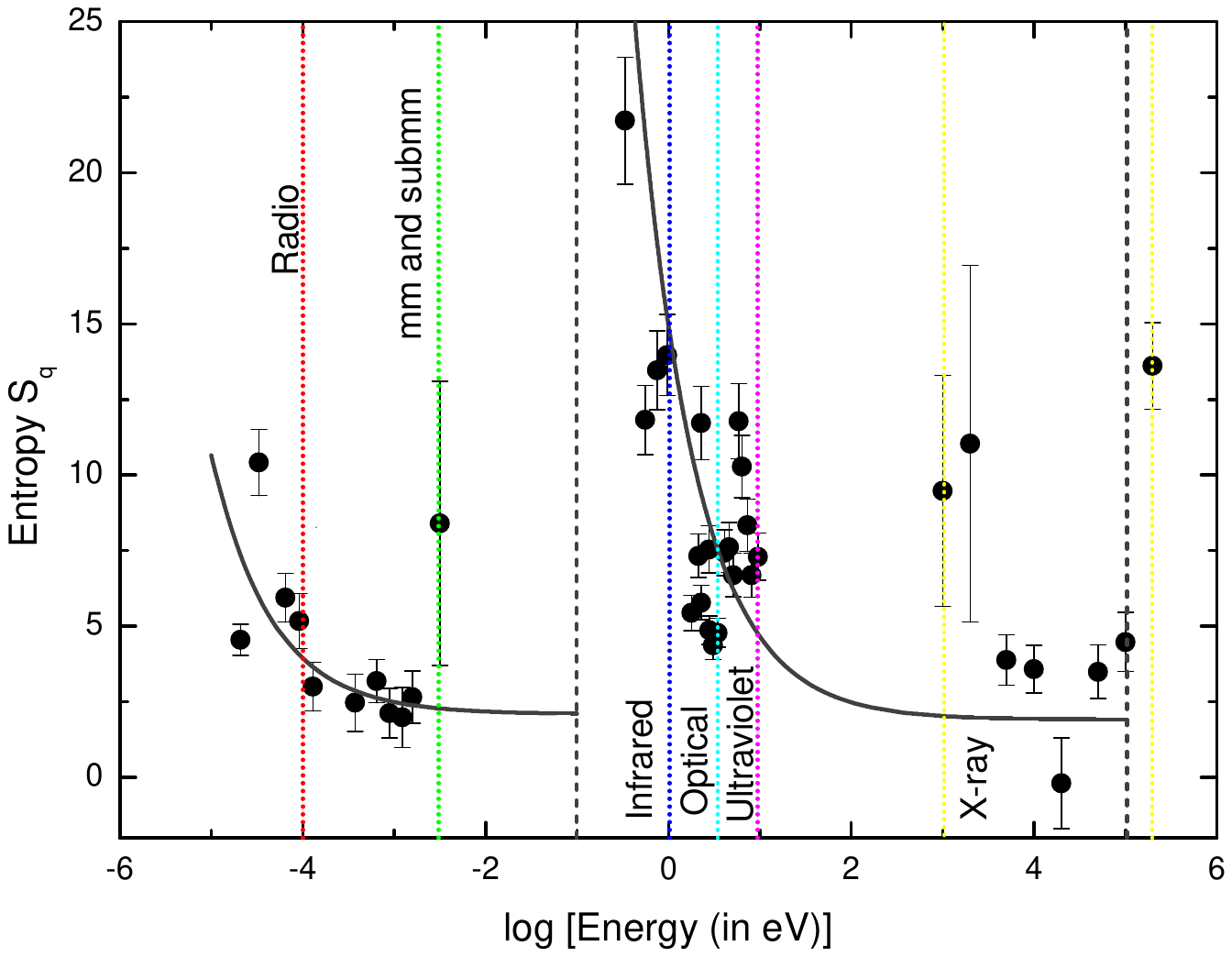}
	\end{center}
\caption{Tsallis entropy values $S_{q}$ as a function of the logarithm of energy band (in eV) for all 39 light curves of Quasar 3C 273. The gray curves denote the best adjustment using the nonlinear L-M method. The black dashed lines denote the asymptotic behavior of non-extensive entropy in a specific energy range. The colored vertical lines represent the upper limit for each band of the electromagnetic spectrum, except for the X-ray band, where the color denotes both the upper and lower limits.}
\label{Fig2}
\end{figure} 

\begin{table}
\scriptsize
\caption[]{\label{tabobs} Observational data selection from the datasets hosted in the ISDC database of Quasar 3C 273.}
\begin{center}
\begin{tabular}{cccc}
 \hline
 \noalign{\smallskip}
 \multicolumn{4}{c}{{\bf ISDC Database}}\\
 \noalign{\smallskip}
 \hline
\noalign{\smallskip}
Spectral Emission & Waveband & Date Range & $\#$ Obs.\\
\noalign{\smallskip}
\hline
\noalign{\smallskip}	
Radio					 & 5 GHz & 1967-2006 & 870\\
 & 8 GHz & 1963-2006 & 1568\\
												 & 15 GHz & 1963-2006 & 1286\\
												 & 22 GHz & 1976-2004 & 1099\\
												 & 37 GHz & 1970-2006 & 1259\\
\hline												
\noalign{\smallskip}
Millimeter				 & 3.3 mm & 1965-2006 & 1548\\
and submillimeter 			 & 2.0 mm & 1981-1997 & 203\\
												 & 1.3 mm & 1981-2007 & 545\\
             & 1.1 mm & 1973-2007 & 322\\
											     & 0.8 mm & 1981-2007 & 496\\			& 0.45 mm & 1982-1996 & 79\\							
\hline
\noalign{\smallskip}
Infrared						 & L(3.6 $\mu$m) & 1969-1997 & 164\\
												 & K(2.2 $\mu$m) & 1967-2004 & 350\\
												 & H(1.65 $\mu$m) & 1967-2004 & 317\\
												 & J(1.25 $\mu$m) & 1976-2004 & 280\\
\hline
\noalign{\smallskip}
Optical 						 & R(7000 \AA) & 1977-2002 & 185\\
& G(5798 \AA) & 1985-2003 & 438\\
												 & V(5479 \AA) & 1968-2005 & 730\\
                                                 & V1(5395 \AA) & 1985-2003 & 438\\
                                                 & B2(4466 \AA) & 1985-2003 & 438\\
												 & B(4213 \AA) & 1968-2005 & 755\\
                                                & B1(4003 \AA) & 1985-2003 & 438\\
												 & U(3439 \AA) & 1968-2005 & 680\\
	
	\hline	
\noalign{\smallskip}		Ultraviolet 					 & 3000 \AA & 1978-2005 & 210\\
& 2700 \AA & 1978-1996 & 199\\
& 2425 \AA & 1978-1996 & 210\\
& 2100 \AA & 1978-2005 & 191\\
												 & 1950 \AA & 1978-1996 & 237\\
												 & 1700 \AA & 1978-1996 & 238\\
             & 1525 \AA & 1978-1996 & 239\\
												 & 1300 \AA & 1978-1996 & 235\\																									\hline	
X-ray 		
& 1 KeV & 1969-2005 & 69\\
& 2 KeV & 1969-2005 & 93\\
& 5 KeV & 1970-2005 & 1032\\
												 & 10 KeV & 1974-2005 & 1026\\
             & 20 KeV & 1976-2005 & 1093\\
												 & 50 KeV & 1977-2005 & 1108\\
												 & 100 KeV & 1978-2005 & 1107\\
             &200 KeV & 1978-2005 & 149\\
\hline			
\end{tabular}
\end{center}
\end{table}

\section{Nonextensive theoretical framework}
The entropy prototype under consideration is the Boltzmann–Gibbs (BG) entropy, a key concept in statistical mechanics and information theory. Over time, this entropy has been generalized into various entropy-like indices, each emerging from distinct scientific frameworks \cite{kol,ren}. Tsallis entropy plays a significant role in out-of-equilibrium statistical mechanics \cite{tsallis1988}. This generalized entropy aims to reconcile the BG entropy with the Khinchin–Shannon axioms, which govern the core properties of entropy \cite{amigo}. However, systems with long-range interactions challenge the fourth axiom, also known as the additivity axiom, which states that the entropy of a composite system $A + B$ should equal the sum of the individual entropies $S$, i.e., $S(A+B) = S(A) + S(B)$.

The BG and Tsallis entropies are two fundamental concepts in statistical mechanics. They differ in their treatment of probability distributions and in the underlying assumptions about the systems they describe. The BG entropy is defined as
\begin{equation}\label{bg1}
S_{BG} = -k_B \sum_{i} P_i \ln P_i,
\end{equation}
where \(k_B\) is the Boltzmann constant and \(P_i\) is the probability of the system being in the \(i\)-th microstate, is additive and assumes that the system is extensive and ergodic \cite{tsallis1988}. In this case, the entropy of a composite system is simply the sum of the entropies of its subsystems, reflecting a classical view of thermodynamics where systems tend to equilibrium.

In contrast, Tsallis entropy is formulated to accommodate non-extensive systems, where interactions between systems can lead to non-additive behavior given by the approach $S(A\oplus B)=S(A)+S(B)+(1-q)S(A)S(B)$, where the additional term represents the interaction between systems $A$ and $B$, which is absent in extensive formalism \cite{tsallis1988,Tsallis3,Tsallis4}. At the core of Tsallis' non-extensive statistical mechanics is the Tsallis entropy. It is defined as
\begin{equation}\label{g4}
        S_{q}=k_{B}\frac{1-\sum_{i} P^{q}_{i}}{q-1},
\end{equation}
where $q$ is a parameter that characterizes the degree of non-extensivity. For $q = 1$, Tsallis entropy reduces to the BG entropy, thereby linking the two frameworks. The non-additive nature of Tsallis entropy allows it to describe systems that exhibit long-range interactions or memory effects, which are often encountered in complex systems such as those found in non-equilibrium statistical mechanics \cite{10.1103/physreve.91.042143}. 

The values of $q$-entropic indexes are derived from probability distribution functions (PDF) named by $q$-Gaussians, $P_{q}$ \cite{viana2020non}. These functions are obtained from the variational problem using the continuous version for the non-extensive entropy given by Eq. \ref{g4} and, therefore, defined by
\begin{equation}\label{g12}
        P_{q}(y)=A_{q}\left[1+(q-1)\left(\frac{y}{\sigma_{q}}\right)^{2}\right]^{1/(1-q)},
    \end{equation}
where $y$ is defined using the condition derived from Pluchino \& Rapisarda \cite{10.1117/12.772041} and, therefore, given by expression $y=(x-\langle x\rangle)/\sigma$. In this case, the PDFs are normalized by subtracting from the $x$'s their average $\langle x\rangle$ and dividing by the correspondent standard deviation $\sigma$, where $x$ denotes the Quasar 3C 273 flux (in Jy). For the parameter $A_{q}$, there are two conditions
    \begin{equation}
        A_{q}=\frac{\Gamma\left(\frac{5-3q}{2-2q}\right)}{\Gamma\left(\frac{2-q}{1-q}\right)}\sqrt{\left(\frac{1-q}{\pi}\right)\beta_{q}}, \quad q<1,
    \end{equation}
and,
    \begin{equation}\label{g14}
        A_{q}=\frac{\Gamma\left(\frac{1}{q-1}\right)}{\Gamma\left(\frac{3-q}{2q-2}\right)}\sqrt{\left(\frac{q-1}{\pi}\right)\beta_{q}}, \quad q>1,
    \end{equation}
    where
    \begin{equation}\label{g15}
        \beta_{q}=\left[(3-q)\sigma^{2}_{q}\right]^{-1}
    \end{equation}
    and
    \begin{equation}\label{g16}
        \sigma^{2}_{q}=\sigma^{2}\left(\frac{5-3q}{3-q}\right),
    \end{equation}
where $\sigma_{q}$ means the generalized standard deviation as a function of $q$, whereas $\sigma$ is the canonical standard deviation.

\section{Results and Discussions}\label{results}

\subsection{Jump-like behavior of Tsallis entropy}
Firstly, Figure \ref{FigqRadio8GHz} is an example that shows that all 39 light curves analyzed obey to Tsallis distributions. This outcome could be explained by the longer tails of distributions that more accurately reflect the intensity variations of Quasar 3C 273 light curves, as suggested by the $q>1$ values. Specifically, for the 8 GHz Radio light curve (see Fig. \ref{FigqRadio8GHz}), the $q$-Gaussian fitting provides us a correlation coefficient of $R^{2}= 0.997$ (indicated by the red line) and $q=1.60$. This adjustment was obtained using a nonlinear regression method based on the Levenberg-Marquardt (L-M) algorithm \cite{83b09f23-b20e-3617-8f72-24765b713f7b,doi:10.1137/0111030}.

Conversely, the Gaussian fitting (denoted by the black dashed line) fails to capture the 8 GHz Radio flux values above $1\sigma$. This pattern persists across all light curves, except in the IR band at 0.34 eV, where a Gaussian profile prevails, accompanied by the lowest $q$-index value in our dataset. For this specific band, we found the maximum value of Tsallis entropy that drives the central inquiry of our work.

Figure \ref{Fig2} reveals a steep increase in the Tsallis entropy in two energy levels of Quasar 3C 273 indicated by the vertical lines. This jump-like behavior is observed in the IR and X-ray spectra but is particularly evident in the far-IR range.  We use the same nonlinear regression method to estimate the best exponential fit, as follows:
\begin{equation}\label{eq1}
S_{q}=A+B\exp\left(-\frac{E}{E_{0}}\right),
\end{equation}
where the parameters $A$ and $B$ for the two domains are $A=2.10\pm 0.92$, $B=0.01\pm 0.023$, and the energy $E_{0}=0.65\pm 0.46$ eV in the range between radio and mm/submm regime, while $A=1.91\pm 2.11$, $B=9.04\pm 2.43$, and $E_{0}=0.64\pm 0.18$ eV are the values for the exponential decay in the range between IR and X-rays with the correlation coefficient stands at 0.51 and 0.75, respectively. To perform the exponential fit, we considered the lower limit of the entropy $S_{q}$ for the three light curves with the largest error bar (those curves with the smallest data).

The jump-like behavior observed in the entropy-energy relationship of Quasar 3C 273 in the IR and X-ray ranges can be attributed to various physical processes occurring in the Quasar. The steep increase in entropy in the IR range followed by an exponential decay until X-rays and another sudden increase in entropy in the X-ray range may be explained by the interplay of different physical mechanisms according to the electromagnetic spectrum range of the Quasar.

\subsection{Physical implications of jump-like increase}

According to Soldi \textit{et al.} \cite{2008A&A...486..411S}, AGNs are highly variable objects at all wavelengths; hence, various approaches can be utilized to measure the magnitude of the fluctuations. Using the same sample as in our work, these authors measured the total variability of each light curve using a parameter defined as the amplitude of the fractional variability ($F_{\rm var}$) that depends on the sample variance, the mean flux and the uncertainties of the measurements (see equation 1 of this paper). They found an increasing trend in the variability $F_{\rm var}$ reaching a peak in the millimeter band around 0.35 mm (see also Figure 4 of this paper). However, they attributed this maximum peak as uncertain due to the scarcity of data in the range of $10^{-2}$ and $10^{-1}$ eV (far-IR range). This fact is also reiterated in our data. From the radio to the submillimeter range, our results indicate that the Tsallis entropy decreases exponentially until reaching a minimum in the far-IR range. On the other hand, the results of Turler \textit{et al.} \cite{1999A&AS..134...89T} and Soldi \textit{et al.} \cite{2008A&A...486..411S} show that in this range $F_{\rm var}$ it grows linearly until reaching a maximum (albeit uncertain) in the far-IR range. However, there is an evident abrupt jump/drop towards the IR range in both cases. In this sense, the parameter $F_{\rm var}$ and the entropy $S_{q}$ are sensitive to this sudden change. We also found that the entropic index has a linear growth up to the far-IR range, a result in agreement with the growth observed from $F_{\rm var}$ by Soldi \textit{et al.} \cite{2008A&A...486..411S}. Since the index $q$ is sensitive to strong fluctuations that widen the tail of the distribution, cumulative effects can trigger a sudden release of energy and, therefore, cause a steep increase in entropy.

According to Turler \textit{et al.} \cite{2000A&A...361..850T} and Soldi \textit{et al.} \cite{2008A&A...486..411S}, the steep increase in variability from radio to millimeter is due to the evolution of synchrotron flares, which are believed to propagate from higher to lower energies. Consequently, consecutive flares overlap, creating an intense and continuous radio emission with relatively small variations. In contrast, very short flares stand out in the millimeter band in a lower quiescent emission, resulting in a large amplitude of fractional variability. As reported by Turler \textit{et al.} \cite{2000A&A...361..850T}, the flares observed in the millimeter band are closely linked to the synchrotron emission from relativistic electrons accelerated in the jet. This phenomenon is consistent with findings from Lindfors \textit{et al.} \cite{10.1051/0004-6361:20053679}, who applied similar methodologies to analyze flaring in another quasar, 3C 279, indicating that the shock-in-jet model effectively describes the outburst dynamics. The synchrotron process is characterized by the emission of radiation from charged particles spiraling in magnetic fields, which is particularly relevant in the context of 3C 273's jet, where the magnetic field strength and particle density can vary significantly during flaring events \cite{10.3847/0004-637x/817/2/121}.

Moreover, the spectral energy distribution of the jet in 3C 273 shows that millimeter flares often correlate with variability in other wavelengths, such as optical and X-ray emissions. This correlation is evident in Fig. \ref{Fig2}, where the entropy shows similar behavior before and after the first black dashed line, indicating a strong relationship between the phenomena occurring in these two spectral regions. This correlation suggests that the emission regions are co-spatial, supporting the notion of a unified emission mechanism across different bands \cite{10.1051/0004-6361:20047021,10.1086/423165}. For example, Jester \textit{et al.} \cite{10.1051/0004-6361:20047021} highlighted that the optical and X-ray emissions are likely dominated by high-energy components rather than the traditional radio synchrotron component, which has implications for understanding the overall energy budget and particle dynamics in the jet.

Additionally, the flaring activity in the millimeter band can be influenced by the underlying physical conditions, such as the viscosity of the emitting plasma and the presence of shock waves. Turler \textit{et al.} \cite{2000A&A...361..850T} noted that variations in these conditions could lead to different spectral characteristics during flares, including the potential for synchrotron self-Compton (SSC) processes to dominate under certain circumstances \cite{10.1093/mnras/stu540}. These processes are particularly relevant when considering the implications for the observed X-ray emissions, which may arise from the same electron population responsible for the millimeter flares \cite{10.1051/0004-6361:20030983}.
Also, as reported by Dai \textit{et al.} \cite{10.1088/0004-637x/753/1/33}, dust-rich environments in quasars may influence the entropy-energy relationship in the infrared range. More recently, Figaredo \textit{et al.} \cite{Figaredo_2020} investigated the characteristics of the dust torus surrounding the Quasar 3C 273. The authors employ reverberation mapping techniques to analyze the time delay between the variations in the luminosity of the central source and the corresponding response from the surrounding dust. Their findings suggest that infrared emission is strongly influenced by thermal radiation from nearby hot dust. This dust exists close to the sublimation radius, where it can withstand the intense radiation from Quasar 3C 273.

Soldi \textit{et al.} \cite{2008A&A...486..411S} demonstrated that synchrotron flares in the IR range (1 to 10 $\mu$m) increase the variability, $F_{\rm var}$, by a factor of 2. Since our data do not rule out the presence of these flares, they strongly influence the entropy drop up to approximately 1 eV. Thus, the presence of these flares suggests that entropy is sensitive to the high-energy spikes associated with them. In summary, according to these authors, synchrotron emission -- extrapolated from the millimeter band -- along with dust, explains the far- and mid-IR emissions, while synchrotron radiation accounts for the emission mechanisms in the near-IR region.

From the optical to the ultraviolet range, $S_{q}$ values show some dispersion and a weak increasing trend ($R^{2}\sim 0.1$), as observed between the vertical cyan and magenta lines in Fig. \ref{Fig2}. In contrast, Soldi \textit{et al.} \cite{2008A&A...486..411S} reported that $F_{\rm var}$ begins to increase from the L band toward shorter wavelengths, reaching up to approximately 10 eV. This trend aligns with the decomposition suggested by Paltani \textit{et al.} \cite{1998A&A...340...47P}, where the red component (R), dominant in the optical range, exhibits smaller variations than the blue component (B), which dominates at UV frequencies. The total emission thus reflects changes in the relative fluxes of these two components. While these explanations may help clarify the scattering of $S_{q}$ in this region, our results only confirm that the overall trend of decreasing entropy remains maintained.

Due to the selection criteria for the light curves of Quasar 3C 273, there is a wide gap between the energies of 10 eV and 1 keV (from UV to soft X-ray). Soldi \textit{et al.} \cite{2008A&A...486..411S} fill this gap filled in the range of 100eV to 1keV, presenting an increasing trend in the behavior of $F_{\rm var}$. However, the curves in this range are extremely sensitive to adding or subtracting a few data points, making the $S_{q}$ measurements highly uncertain in this band. Differences in X-ray spectra between radio-quiet and radio-loud quasars, as highlighted in studies \cite{10.1046/j.1365-8711.2000.03510.x} comparing a large sample of quasars, could play a role in this decay of entropy towards the X-ray range.

For some hard X-ray and $\gamma$-radiation light curves from the HEAVENS interface, the entropy $S_{q}$ cannot be calculated because of the small amount of available data or the large measurement uncertainties (0.5–10 MeV and 3–10 GeV). Consequently, the increase of $S_{q}$ observed in the X-rays (second vertical black line from Fig. \ref{Fig2}) is most likely due to large flux uncertainties. In addition, as mentioned by Soldi \textit{et al.} \cite{2008A&A...486..411S}, caution should be used when considering the variability results in the $\gamma$-ray band above 30 MeV because these light curves also include fluxes obtained from the same observations by different analyses.

Another way to analyze whether the second peak is real is to analyze the correlations between the millimeter band and the X-ray radiation. This hypothesis is based, as mentioned above, on the same type of exponential decay that separates the first peak at 0.34eV. According to Soldi \textit{et al.} \cite{2008A&A...486..411S}, a lack of short-term correlation between the mm and the X-ray radiation indicates no clear connection between the X-rays and the mm radio flare emission. Thus, as suggested by Marscher \& Gear \cite{1985ApJ...298..114M}, if the electron cooling process, which dominates the first phase of jet shock evolution, were producing most of the X-ray emission, the mm flares would follow the X-ray flares with small delays. However, Soldi \textit{et al.} \cite{2008A&A...486..411S} did not observe such a correlation, suggesting that the simultaneous emission of both the mid- and IR and X-rays may result from a secondary process. This lack of correlation between the millimeter and X-ray bands shows that some synchrotron events can be so energetic that they can surpass the X-ray emission even at energies above a few tens of keV.

In summary, only the first peak in the far-IR has physical implications, evidenced mainly by the dynamics of the flares. In contrast, the second peak in the hard X-ray range towards the  $\gamma$-rays has no statistical relevance. Consequently, more detailed studies using data from other sources, such as CGRO/COMPTEL and EGRET data, and more robust statistical procedures to treat light curves with few data will be useful to investigate whether the second peak is real or a statistical artifact.


\section{Final remarks}
Our study showed that the light curves of Quasar 3C 273 predominantly follow heavy-tailed Tsallis distributions. The estimated $q$ values from $q$-Gaussian distributions showed high correlation coefficients, outperforming the traditional Gaussian fitting. This pattern was consistent across wavelengths, except for the IR band at 0.34 eV, where a Gaussian profile performed the best fit. Coincidentally, the Tsallis entropy has its maximum value for this energy value.

In addition, the entropy-energy relationship in Quasar 3C 273 exhibited a jump-like behavior, particularly in the IR and hard X-ray spectra. The entropy reached its first peak in the IR range and then decayed exponentially until reaching a second peak in the hard X-ray spectrum (no statistical relevance). This jump-like behavior in the IR range was attributed to different emission mechanisms, mainly due to the dynamics of the flares and dust.

The increased entropy in the IR range has been linked to processes such as synchrotron radiation and dust-rich environments around the quasar 3C 273. The subsequent exponential decay towards X-rays is associated with changes in the emission mechanisms occurring in the optical-UV range. The increase in entropy at X-ray wavelengths is uncertain due to the poor quality of the light curves.

The results obtained by comparing the Quasar 3C 273 data and the non-extensive statistical framework show promise in other astrophysical scenarios involving active galactic nuclei, such as cataclysmic variables, pulsars, BL Lacertae objects, supernova remnants, and other X-ray sources. In this sense, it is crucial to investigate whether the jump-like behavior found in Quasar 3C 273 is also observed in other astrophysical sources or is just a phenomenon particular to this Quasar. On the other hand, since all these sources manifest diffusive processes in which particles in different energy ranges are submitted to their strong gravitational fields, they are expected to behave similarly concerning the Tsallis distribution.

\acknowledgments
DBdeF acknowledges financial support from the Brazilian agency CNPq-PQ2 (Grant No. 305566/2021-0). Research activities of STELLAR TEAM of Federal University of Cear\'a are supported by continuous grants from the Brazilian agency CNPq.


\bibliographystyle{eplbib}
\bibliography{3C273EPL2025}

\end{document}